\documentclass[twocolumn,showpacs,floatfix,aps]{revtex4}

\def \beq {\begin{equation}}
\def \eeq {\end{equation}}

\usepackage[dvips]{graphics}

\begin{document}

\title{A note on a third order curvature invariant in static spacetimes}

\author{Alberto Saa}
\email{asaa@ime.unicamp.br}
\affiliation{
Departamento de Matem\'atica Aplicada, \\
IMECC -- UNICAMP, 
C.P. 6065,\\ 13083-859 Campinas, SP, Brazil.}


\begin{abstract}
We consider here the third order curvature invariant 
$I=R_{\mu\nu\rho\sigma;\delta}R^{\mu\nu\rho\sigma;\delta}$
in static  spacetimes ${\cal M}=R\times\Sigma$ for which $\Sigma$ is
conformally flat. We evaluate explicitly the invariant   for
the  $N$-dimensional    Majumdar-Papapetrou multi
black-holes solution, confirming that 
$I$  does indeed vanish on the 
event horizons of such black-holes.  
Our calculations show, however,   that solely the 
vanishing of $I$  is not sufficient to locate an event horizon in  
non-spherically symmetric spacetimes. We discuss also 
some tidal effects associated to 
the invariant  $I$.
\end{abstract}

\maketitle

Recently, the third order curvature invariant 
$I=R_{\mu\nu\rho\sigma;\delta}R^{\mu\nu\rho\sigma;\delta}$
has received some attention in the literature. The observation that $I$ could be
used to single out the event horizon in the Schwarzschild spacetime can be traced back
to \cite{rn}. It is not difficult to show that, for spherically symmetric static
black-holes, $I$ is positive in the exterior region  and vanishes
on the black-hole event horizon. Thus, in principle,
some specific local measurements\cite{t} could be indeed employed
 by in-falling observers to detect the crossing of the event horizon of spherically
 symmetric black-holes.
 Several other aspects and properties of higher order  curvature invariants
 have been also examined\cite{a1,con,a2,G,G1}. 

In \cite{jhep}, the invariant $I$ is considered for static 4-dimensional Einstein 
spacetimes  ${\cal M}=R\times\Sigma$, with $\Sigma$   conformally flat. 
Several properties of the invariant and some relations to the topology
of $\Sigma$ are discussed. Here, we investigate 
the 
invariant $I$   for static  $N$-dimensional  spacetimes ${\cal M}=R\times\Sigma$
with $\Sigma$  conformally flat, but without any further 
assumptions on $\cal M$.
 Since $\cal M$ is assumed to be static, its metric
can be cast in the form 
\beq
\label{metric}
ds^2 = g_{\mu\nu}dx^\mu dx^\nu = -f^2dt^2 + h_{ij} dx^idx^j ,
\eeq
where $h_{ij}$ is the $(N-1)$-dimensional Riemannian metric of $\Sigma$
and  $f$ is 
 smooth function on $\Sigma$. Greek indices run over 0 to $N-1$, whereas Latin ones are reserved
to the spatial coordinates of $\Sigma$  and run, unless specified otherwise, over 1 to $N-1$.

The non-vanishing components of
the Riemann tensor of the metric (\ref{metric}) are
\beq
R_{0i0j} = f \hat\nabla_i\hat\nabla_j f, \quad
R_{ijkl} = \hat R_{ijkl}.
\eeq
The hat here  denotes      intrinsic quantities of $\Sigma$. 
The covariant derivative of the Riemann tensor can be also evaluated
\begin{eqnarray}
\label{dr}
R_{0i0j;k} &=&   f\hat\nabla_k\hat\nabla_i\hat\nabla_j f - (\hat\nabla_k f)\hat\nabla_i\hat\nabla_jf, \nonumber \\
R_{0ijk;0} &=&  (\hat\nabla_i\hat\nabla_j f)\hat\nabla_k f - (\hat\nabla_i\hat\nabla_k f)\hat\nabla_j f
- f\hat R_{lijk}\hat\nabla^lf, \nonumber \\
R_{ijkl;m} &=& \hat R_{ijkl;m},
\end{eqnarray}
leading to
\beq
\label{I}
I = 4 R_{0ijk;0}R^{0ijk;0} + 4R_{0i0j;k}R^{0i0j;k} + \hat R_{ijkl;m}\hat R^{ijkl;m}.
\eeq
Note  that each term in (\ref{I}) is non-negative for   metrics of the form (\ref{metric}).
From the assumption of a conformally flat $\Sigma$, one can choose a coordinate system on
$\Sigma$ such that $h_{ij}=h^2\eta_{ij}$, where $h$ is a smooth function and $\eta_{ij}$ is
a flat $(N-1)$-dimensional metric.

\begin{figure}[ht]
\resizebox{1.3\linewidth}{!}{\includegraphics*{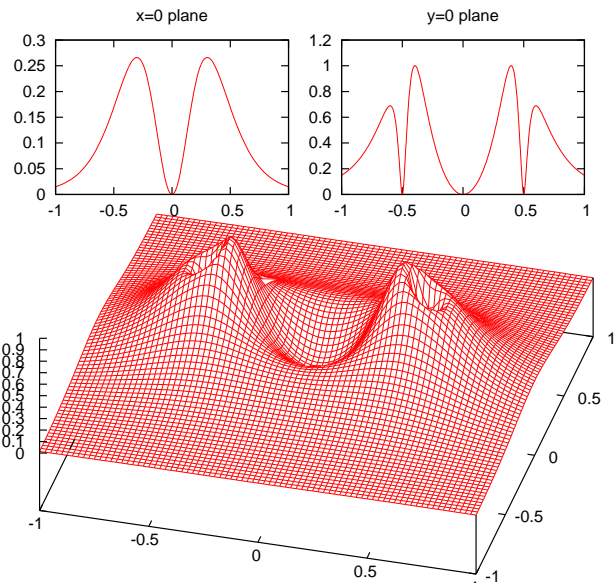}}
\caption{The invariant $I=R_{\mu\nu\rho\sigma;\delta}R^{\mu\nu\rho\sigma;\delta}$
for the two equal masses Majumdar-Papapetrou black-holes described by (\ref{2bh}), with $M=2a=1$.
All the invariants of this work were calculated numerically 
by simple central differences, with good accuracy  and little
computational efforts\cite{Saa}.}
\label{figure1}
\end{figure}

\begin{figure}[ht]
\resizebox{1.3\linewidth}{!}{\includegraphics*{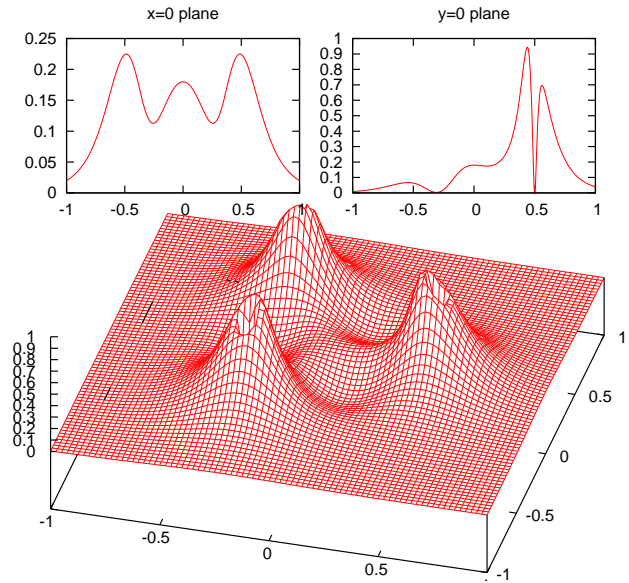}}
\caption{The invariant $I=R_{\mu\nu\rho\sigma;\delta}R^{\mu\nu\rho\sigma;\delta}$ 
for three unit mass Majumdar-Papapetrou black-holes placed on the vertices of
an equilateral triangle inscribed in a circle with radius $1/2$.}
\label{figure2}
\end{figure}

Our first observation is that $I$ also vanishes  on the  horizons of the
 $N$-dimensional Majumdar-Papapetrou multi black-holes  solution. Such a solution (see \cite{mbh}
for further references) 
correspond to the choice  $f = U^{-1}$ and $h = U^\frac{1}{N-3}$, with
\beq
U = 1 + \sum_a \frac{m_a}{|X-X_{(a)} |^{N-3}},
\eeq
where $m_a$ stands for the mass of the (extremal) charged black-hole placed at the
point $X_{(a)}\in\Sigma$. The horizons of the Majumdar-Papapetrou solution are located precisely 
at the points $X_{(a)}$. 
Analogously to the 4-dimensional case\cite{hawking},
such horizons do indeed correspond to hypersurfaces of
$\Sigma$ with area $A_{N-2} m_a^{2/(N-3)}$,  
where $A_{N-2}$ stands for the area of the unit $(N-2)$-dimensional 
hypersphere.
They were shrunk to single points here only as a
consequence of the 
choice of the coordinate system (\ref{metric}). In order to show that $I$ vanishes
on a given horizon $X_{(a)}$, 
let us introduce $(N-1)$-dimensional 
spherical coordinates $(r,\theta_1,\dots,\theta_{N-2})$
centered in $X_{(a)}$ and consider small $r$, leading to
$ U \approx {m_a}/{r^{N-3}}$ and, consequently, to 
\beq
f = \frac{r^{N-3}}{m_a},\quad h = \frac{m_a^\frac{1}{N-3}}{r}.
\eeq
By introducing the local orthonormal frame 
\begin{eqnarray}
\label{of}
\omega^{\hat{t}}&=& fdt, \quad \omega^{\hat{r}} = hdr,  \\
\omega^{\hat{\theta}_i} &=& hr\left(\prod_{1\le j <i}\sin\theta_j\right)d\theta_i, \quad i=1\dots N-2,\nonumber
\end{eqnarray}
the non-vanishing components of the 
Riemann tensor around a given horizon $X_{(a)}$ of the $N$-dimensional Majumdar-Papapetrou metric read
simply
\begin{eqnarray}
\label{r}
R_{\hat t\hat r\hat t\hat r} &=& -(N-3)^2m_a^{-\frac{2}{N-3}},\nonumber\\
R_{\hat \theta_i\hat \theta_j \hat \theta_i\hat \theta_j} &=& m_a^{-\frac{2}{N-3}},\quad (i\ne j). 
\end{eqnarray}
From (\ref{r}), one can show that the first covariant derivative  of 
the Riemann tensor vanishes in a close
neighborhood of a Majumdar-Papapetrou
horizon, implying, of course, the vanishing of the invariant $I$.
 However, in contrast 
with the spherically symmetric case\cite{con},  $I$ cannot be
used as  a ``horizon detector" for the multi-black holes solution  
since it also vanishes
for other points with no relation with   horizons. Let us illustrate this fact with some
explicit 4-dimensional situations (See Figs. \ref{figure1}  and \ref{figure2}).
The case of two   equal masses black-holes  separated by a distance of $2a$
corresponds to the the choice 
\beq
\label{2bh}
U = 1 +  \frac{M}{\sqrt{(x+a)^2+y^2+z^2}} + \frac{M}{\sqrt{(x-a)^2+y^2+z^2}}.  
\eeq
The invariant $I$ for such configuration is depicted in Fig. \ref{figure1}.
Due to the symmetry of $U$ under the  total reflection 
$(x,y,z)\rightarrow(-x,-y,-z)$ and the smoothness of the multi black-hole solution, all 
odd-order derivatives of the Riemann tensor  
should vanish in the origin,
implying, in particular, that $I(0,0,0)=0$, as also illustrated in Fig. \ref{figure1}.
Fig. \ref{figure2} depicts an equilateral configuration of three equal masses
black-holes.

Before starting the discussion of the tidal effects associated to the invariant
$I$, we notice also that $I$ should vanish for any isolated event horizon of metrics like
(\ref{metric}), not only for the
Majumdar-Papapetrou case. In order to show that, let us consider again the
$(N-1)$-dimensional 
spherical coordinates around a given generic isolated horizon. Without
loss of generality, the metric near the horizon can be cast in the general
spherically symmetrical form
\begin{equation}
\label{metric2}
ds^2 = - f(r)dt^2 + \frac{dr^2}{f(r)} + e^{\rho(r)}d\Omega^2_{N-2},
\end{equation}
where $f(r)$ and $\rho(r)$ are arbitrary functions of $r$ with
$f(0)=0$.
We make the following assumptions for the metric (\ref{metric2}) 
at the horizon $r=0$:
\begin{enumerate}
\item The volume element $\sqrt{-g} = e^{\rho(r)}\sin\theta$ is smooth and
non vanishing;
\item All curvature invariants are smooth and bounded;
\item The function $f(r)$ obeys
\begin{equation}
\label{power}
f(r) = cr^s + {\cal O} \left(r^{s+1}\right),
\end{equation}
with both c and s positives.
\end{enumerate}
Assumptions 1 and 2 assure the regularity of the horizon. The last
requirement seems general enough to include all physically relevant
horizons.
From the assumptions 1 and 2 one has that $s=1$ or
$s\ge 2$ in (\ref{power}),
as one can verify by evaluating, for instance, the simplest
scalar invariant for the metric (\ref{metric2}), namely its
scalar curvature
\begin{equation}
R = \frac{2}{e^{\rho(r)}} - cs(s-1)r^{s-2} + {\cal O}
\left( 
r^{s-1}
\right).
\end{equation}
We notice that
exactly the same restriction on $s$ is obtained by requiring
a bounded Kretschmann invariant 
$K=R_{\mu\nu\rho\sigma}R^{\mu\nu\rho\sigma}$ at $r=0$. 
The invariant $I$ near the horizon reads
\begin{eqnarray}
\label{II}
I &=& cr^s\left[ \frac{8{\rho'}^2}{e^{2\rho}} + 
 c^2 s^2(s-1)^2(s-2)^2r^{2(s-3)} \right. \nonumber \\
 &+& 2{\rho'}^2c^2s^2(s-1)^2r^{2(s-2)}
+ 4\rho''\rho'c^2s^2(s-1)r^{2s-3} \nonumber \\
 &+& \left. \rule{0cm}{0.6cm} {\cal O} \left( 
r^{2(s-1)} \right)
\right].
\end{eqnarray}
Finally, if
the surface corresponding to $r=0$ 
is a horizon, $s=1$ or $s\ge 2$, and $e^{\rho(r)}$
is a smooth non-vanishing function at $r=0$,
implying, from (\ref{II}), that $I$ vanishes there.

In order to grasp some of the physical meaning of the invariant $I$, 
let us remember 
that for a 4-dimensional Schwarzschild black hole with mass $m$, it reads
\beq
I = 720\left(1-\frac{2m}{r} \right)\frac{m^2}{r^8}.
\eeq
As one can see, it is smooth for any $r>0$, it is
 positive in the exterior region,  
negative in the interior, 
and 
vanishes only for $r=2m$. Furthermore, $I$ attains its maximum value for $r=9m/4$,
it is a monotone increasing function for $0< r < 9m/4$, and a monotone decreasing one
for $r> 9m/4$.  Such a behavior is in contrast with the Kretschmann scalar
for the Schwarzschild black hole,
$K=48m^2/r^6$, which is a monotone decreasing function for all $r>0$.
We notice that 
the radius $r=9m/4$ plays an important role in Schwarzschild spacetimes. It corresponds to
the minimal possible radius for a stable star of mass $m$, {\em i.e.}
if a spherically symmetric body of mass $m$  
has a radius $r<9m/4$, it core pressure diverges and it will unavoidably
collapse into a black-hole\cite{wald}.

The Riemann tensor and its covariant derivatives are related to tidal effects.
For the Schwarzschild case, the non-vanishing components of the Riemann tensor
in the local
orthonormal frame given by
$\omega^{\hat{t}}= \left(1-2m/r \right)^{1/2}dt$, $\omega^{\hat{r}} = dr/\left(1-2m/r \right)^{1/2} $,
$\omega^{\hat{\theta}} = rd\theta$, and $\omega^{\hat{\phi}} = r\sin\theta d\phi$,
read
\begin{eqnarray}
R_{\hat t\hat r\hat t\hat r} = -R_{\hat \theta\hat \phi\hat \theta\hat \phi}  = 
2R_{\hat t\hat \theta\hat t\hat \theta} = 2R_{\hat t\hat \phi\hat t\hat \phi} = \nonumber \\
-2R_{\hat r\hat \theta\hat r\hat \theta} = -2R_{\hat r\hat \phi\hat r\hat \phi} =  
-\frac{2m}{r^3},
\end{eqnarray}
from where we can easily recognize the usual (Newtonian) tidal force $F_{\rm T} = 2m/r^3$. The derivatives
of the Riemann tensor are naturally related to higher order tidal effects, {\em i.e.}
they give origin to non-linear corrections to the tidal force or, equivalently, to
non-quadratic terms in the tidal potential.
The radius $r=9m/4$, as the maximum of  
$I=R_{\mu\nu\rho\sigma;\delta}R^{\mu\nu\rho\sigma;\delta}$,
should correspond also to the boundary of regions where certain
components of 
$R_{\mu\nu\rho\sigma;\delta\gamma}$ have different signs, as illustrated,
for instance, by 
  the components
\begin{eqnarray}
\label{ht}
R_{\hat t\hat r\hat t\hat r;\hat r\hat r} = -R_{\hat \theta\hat \phi\hat \theta\hat \phi;\hat r\hat r}  = 
2R_{\hat t\hat \theta\hat t\hat \theta;\hat r\hat r} = 2R_{\hat t\hat \phi\hat t\hat \phi;\hat r\hat r} = \nonumber \\
-2R_{\hat r\hat \theta\hat r\hat \theta;\hat r\hat r} = -2R_{\hat r\hat \phi\hat r\hat \phi;\hat r\hat r} =  
-24\left(1-\frac{9}{4}\frac{m}{r}\right)\frac{m}{r^5}.
\end{eqnarray}
Far from the Schwarzschild horizon ($r\gg 2m$), we recover from (\ref{ht}) 
the usual higher order Newtonian tidal correction corresponding 
to $\partial^2F_{\rm T}/\partial r^2$. The situation, however, is dramatically
different near the horizon. In particular, inside the radius $r=9m/4$,
the higher order tidal force is indeed attractive!

\begin{figure}[ht]
\resizebox{1.3\linewidth}{!}{\includegraphics*{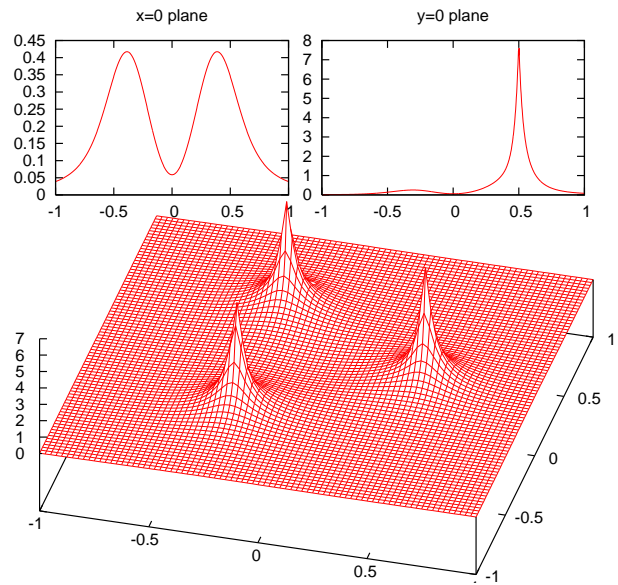}}
\caption{The Kretschmann invariant $K=R_{\mu\nu\rho\sigma}R^{\mu\nu\rho\sigma}$
for the situation depicted in Fig. \ref{figure2}:
 three unit mass Majumdar-Papapetrou black-holes placed on the vertices of
an equilateral triangle. The equivalent of the situation depicted in
Fig. \ref{figure1} is analogous.}
\label{figure4}
\end{figure}

We can see from Figs. \ref{figure1}, \ref{figure2} and \ref{figure4}
that essentially
the same 
behavior near the horizon   holds also for   multi-black holes configurations.
Each Majumdar-Papapetrou horizon is placed in  the hollow bottom of $I$,
surrounded by a ``barrier" with high corresponding to the vanishing of certain
directional derivatives of $I$, whereas $K$ always  decreases as one departs from the
horizon. Analogously to the Schwarzschild case, in the regions corresponding to the
hollows of $I$, one should also expect attractive  higher order tidal forces.

If the higher order tidal effects associated to the derivatives of the Riemann
tensor are in fact measurable in realistic situations is a question
for which we do not have an answer yet. However, as we have shown, one cannot locate
horizons in the non-spherically symmetrical case by only searching for points where
the invariant $I=R_{\mu\nu\rho\sigma;\delta}R^{\mu\nu\rho\sigma;\delta}$ vanishes.

\acknowledgements

The author is grateful to CNPq and FAPESP for the financial support and to G. Matsas
for useful conversations.

\end{document}